\documentclass[aps,prl,twocolumn]{revtex4-1}

\usepackage{amsmath}
\usepackage{amssymb}
\usepackage{natbib}
\usepackage{graphicx}
\usepackage{bm,hyperref}

\def\be{\begin{equation}}
\def\ee{{\bm e}}
\def\ba{\begin{eqnarray}}
\def\ea{\end{eqnarray}}
\def\rr{{\bf r}}

\def\kk{{\bm k}}

\def\e{\epsilon}
\def\ve{\varepsilon}

\def\p{\partial}
\def\l{\ell}
\def\jj{\bm{j}}
\def\pp{{\bm p }}
\def\rr{\bm{r}}
\def\qq{\bm{q}}

\def\EE{\bm{E}}

\def\d{\delta}

\def\bra{\langle}
\def\ket{\rangle}

\begin{document}
\title{Two-particle collisional coordinate shifts and hydrodynamic anomalous Hall effect in systems without Lorentz invariance}

\date{\today}
\author{ D. A. Pesin}
\affiliation{Department of Physics and Astronomy, University of Utah, Salt Lake City, UT 84112 USA}
\begin{abstract}
We show that electrons undergoing a two-particle collision in a crystal experience a coordinate shift that depends on their single-particle Bloch wave functions, and derive a gauge-invariant expression for such shift, valid for arbitrary band structures, and arbitrary two-particle interaction potentials. As an application of the theory, we consider two-particle coordinate shifts for Weyl fermions in space of three spatial dimensions. We demonstrate that such shifts in general  contribute to the anomalous Hall conductivity of a clean electron liquid.
\end{abstract}
\maketitle

\textit{Introduction -- }Studies of quantum corrections to semiclassical transport in itinerant systems have become an interdisciplinary research direction, encompassing the phenomena of anomalous Hall effect\cite{NagaosaReview}, the chiral anomaly in the particle physics\cite{AdlerAnomaly,BellJackiwAnomaly} and condensed matter physics\cite{Nielsen1983, Son2013a}contexts, static and dynamic chiral magnetic effect in quark-gluon plasma\cite{Vilenkin,kharzeevZhitnitsky2007,sonsurowka,schafer2009, KharzeevSon2011,SonYamamoto,Gorbar2016} and metals with nontrivial band geometry\cite{ChenBurkov2013,ChangYang2015,ChangNoWeyl,MaPesin2015,Zhong2016,Alavirad2016,Kharzeev2016,Beenakker2016,Rou2017,Cheianov2017,Rou2017Te}. Most of the studies on the above subjects focused on the single-particle properties of the systems of interest, exploring the effects of band geometry on observable properties (see Ref.~\onlinecite{XiaoNiu} for a comprehensive review of quantum corrections to the semiclassical dynamics in crystals). There are several exceptions to this rule, for instance Refs.~\onlinecite{Chen2014,Chen2015}, which considered how geometry of single-particle wave functions affected two-particle collisions of chiral fermions interacting via a local interaction in free space. Naturally, these considerations relied heavily on the Lorentz (and hence rotational) invariance present in the problem.

In this paper, we consider coordinate shifts of two colliding electrons -- \emph{two-particle coordinate shift} -- in a generic crystalline band structure, interacting via a generic two-particle potential, without the convenience of rotational, or full Lorentz symmetries. The problem is motivated by advances in materials physics, which have yielded itinerant electronic systems of purity level sufficient for the electronic liquid to demonstrate hydrodynamic behavior~\cite{Molenkamp1995,Bandurin2016,Crossno2016,Moll2016,Kumar2017}. These experimental advances triggered theoretical studies of electronic flow in the hydrodynamic regime~\cite{Gurzhi1981,Levitov2016,Polini2016,Levitov2017}.

In the hydrodynamic regime, the transport properties of the electronic liquid are determined by the collisions between carriers, rather than those with phonons or impurities. Therefore, the study of how the quantum mechanics of electrons in crystals affects such collisions, and manifests itself in the hydrodynamic properties, is of fundamental importance. In this work, we consider how the quantum mechanical effects, in particular the geometry of Bloch functions in the Brillouin zone, manifests itself in the anomalous transport properties: the hydrodynamic anomalous Hall effect.

It is well known that one-particle coordinate shifts in electron-impurity and electron-phonon collisions are related to the band geometry of the material~\cite{Sinitsyn2006}, and  play an important role in the transport and optical properties of crystalline materials (see Refs.~\onlinecite{SturmanReview,SinitsynReview} for a review). While the early works on this subject were carried out several decades ago\cite{Luttinger1958,berger1970,Belinicher1982}, the modern band theory of the side-jump process was formulated fairly recently\cite{Sinitsyn2006}. However, to the best of our knowledge, collisional two-particle shifts in crystals have not been yet considered. Below we will show that two-particle collisional coordinate shifts are a new type of quantum-mechanical correction to the semiclassical electron dynamics, distinct from the single-particle ones for indistinguishable particles, and that they make a contribution to the anomalous Hall effect in the hydrodynamic regime.

\textit{Two-particle coordinate shift --} Consider a collision of two electrons in a crystal. Physically, the collision should be thought of as that of two electronic wave packets, centered around certain quasi-momenta, which belong to some bands in the electronic band structure. The location of a single electron in the unit cell of a crystal in general depends on its quasimomentum, hence the change of the quasimomenta upon the collision will lead to a redistribution of the wave packets in the unit cell: the ``center of mass'' coordinates of the two colliding particles shift upon a collision. In the case of collisions of distinguishable particles -- \textit{e.g.} electrons with impurities, or electrons with phonons -- one can trace the initial and final quasimomenta of the colliding electron, and define the corresponding coordinate shift of the electron, which, for weak centrosymmetric impurity potential depends only on the initial and final Bloch states.

In the case of a two-electron collision, the preceding logic applies inasmuch as that the collision must be accompanied by some coordinate shift. However, particle indistinguishability makes it impossible to decide which initial state the electrons in the final states came from, and it is clear that the total displacement of the two-electron system -- the two-particle coordinate shift -- cannot be reduced to any combination of single-particle shifts. In fact, individual coordinate shifts are not defined under such circumstances in general, in the sense that there is no way to express them within the second quantization formalism.

To describe such two-particle coordinate shifts, we define the Hamiltonian, $\hat H$, of electrons in a crystal as $\hat H=\hat H_0+\hat V$, where the first and second terms are the Hamiltonians of non-interacting Bloch electrons, and the electron-electron interaction, respectively. The single-particle eigenstates of $\hat H_0$ are described by Bloch functions,$\psi_{n\pp}$, that have the usual form of
\begin{align}
  \psi_{n\pp}(\rr)=\frac{1}{\sqrt{N}}e^{i\pp\rr}u_{n\pp}(r),
\end{align}
where $u_{n\pp}(r)$ is a spinor  periodic with respect to lattice translations, $N$ is the number of unit cells in the crystal, and we set $\hbar=1$, which will be assumed from here on. The corresponding single-particle energies will be denoted with $\ve_{n\pp}$. In what follows we assume that the spin degeneracy is lifted in the band structure, since the main applications of the theory, \textit{e.g.} Weyl fermions physics, or anomalous Hall effect, all correspond to broken time-reversal, or inversion symmetries in materials with strong spin-orbit coupling.

For notational convenience, we will combine the band index of the Blosh state, $n$, and its quasimomentum, $\pp$, into a single index $(n,\pp)$, which will be denoted with indices $i$ or $f$ whenever it is necessary to emphasize whether the state is one the initial or final scattering states, or index $\ell$ if such identification is not important. In the basis of Bloch eigenstates, the single-particle and interaction parts of the Hamiltonian are given by
\begin{align}\label{eq:Hamiltonian}
  \hat H_0&=\sum_{\l}\ve_{\ell}a^\dagger_{\ell}a_{\ell},\nonumber\\
  \hat V&=\frac{1}{2}\sum_{\l_1\l_2\l_3\l_4}v_{\l_1\l_2;\l_3\l_4}a^\dagger_{\ell_1}a^\dagger_{\ell_2}a_{\ell_3}a_{\ell_4}.
\end{align}
In this work, we assume the absence of Umklapp processes, which implies that the matrix element of the interaction Hamiltonian contains a momentum-conserving factor $\d_{\pp_1+\pp_2,\pp_3+\pp_4}$.

In what follows, we consider collisions of electrons described by Hamiltonian~\eqref{eq:Hamiltonian}, assuming that $\hat V$ can be treated in the Born approximation. Neglecting the Fermi liquid effects allows to derive the coordinate shift for the two electrons considering the collision in vacuum. Indeed, we will see that the shift is determined only by the single-particle wave functions of the colliding electrons, while the presence of other electrons can only lead to Pauli blocking of the collision itself, or provide RPA-type renormalization of the interaction matrix element. Both effects do not bring any essential physics into the consideration of the coordinate shift.

The process of semiclassical scattering of two electrons described by Hamiltonian~\eqref{eq:Hamiltonian} can be visualized as the motion of two wave packets, with their momenta centered around the initial states $i_1=(\kk_1,n_1)$ and $i_2=(\kk_2,n_2)$ at time $t\to-\infty$, which scatter into wave packets with momenta centered around states $f_1=(\pp_1,m_1)$ and $f_2=(\pp_2,m_2)$ at $t\to+\infty$. The two-particle state corresponding to the incoming pair of particles can be written as
\begin{align}\label{eq:wavepacket}
  |\psi^{\textrm{in}}\rangle =\sum_{\qq_1,\qq_2}w(\qq_1,\qq_2)a^\dagger_{\ell_1} a^\dagger_{\ell_2}|0\rangle,
\end{align}
with $\ell_1=(\qq_1,n_1)$, and $\ell_2=(\qq_2,n_2)$, and the antisymmetric, $w(\qq_1,\qq_2)=-w(\qq_2,\qq_1)$, wave packet amplitude restricts $\qq_1$ and $\qq_2$ to the vicinity of quasimomenta $\kk_1$ and $\kk_2$ in bands $n_1$ and $n_2$. For well-separated in momentum space electrons in the initial state, the wave packet amplitude can be written as $w(\qq_1,\qq_2)=(w_{\kk_1}(\qq_1)w_{\kk_2}(\qq_2)-w_{\kk_1}(\qq_2)w_{\kk_2}(\qq_1))/\sqrt{2}$, where individual amplitudes $w_{\kk}(\qq)$ are functions of $\qq$ centered around $\kk$, and normalized according to $|w_\kk(\qq)|^2=1$.

The evolution of wave packet~\eqref{eq:wavepacket} under the action of the interaction Hamiltonian, $\hat V$ of Eq.~\eqref{eq:Hamiltonian}, is described by standard quantum mechanics, and is analogous to the considerations of Ref.~\cite{Sinitsyn2006}. The details will be presented elsewhere~\cite{footnoteAHE}. Here we only mention that the expectation value of the ``total coordinate'' $\bm R=\rr_1+\rr_2$ of the colliding electrons before and after the collision can be written as
\begin{align}\label{eq:deltardefined}
  \bm R(t\to-\infty)&=\bm v_{i_1}t+\bm v_{i_2}t+\rr^{\textrm{in}},\nonumber\\
   \,\,\bm R(t\to+\infty)&=\bm v_{f_1}t+\bm v_{f_2}t+\rr^{\textrm{out}},\nonumber
\end{align}
where the group velocities of electrons in the initial and final states are denoted with $\bm v$ with appropriate subscripts. The two-particle collisional coordinate shift is defined as
\begin{align}
   \d \rr^{f_1f_2}_{i_1i_2}&=\rr^{\textrm{out}}-\rr^{\textrm{in}},
\end{align}
for which we obtain~\cite{footnoteAHE}
\begin{widetext}
\begin{align}\label{eq:shiftfinal}
   \d \rr^{f_1f_2}_{i_1i_2}&=
   i\bra u_{f_1}|\p_{\pp_1}u_{f_1}\ket+i\bra u_{f_2}|\p_{\pp_2}u_{f_2}\ket
   -i\bra u_{i_1}|\p_{\kk_1}u_{i_1}\ket-i\bra u_{i_2}|\p_{\kk_2}u_{i_2}\ket
   -(\p_{\pp_1}+\p_{\pp_2}+\p_{\kk_1}+\p_{\kk_2})\arg V_{f_1f_2;i_1i_2}.
 \end{align}
\end{widetext}
 The matrix element of the interaction Hamiltonian is given by $V_{f_1f_2;i_1i_2}=\langle f_1f_2| \hat V|i_1i_2\rangle$. Eq.~\eqref{eq:shiftfinal} is the central result of this work; it provides a gauge-invariant expression for the two-particle coordinate shift, in the sense that it does not depend on the momentum-dependent phase choice for the Bloch wave functions.

It is straightforward to show that for distinguishable particles the two-particle shift reduces to the sum of the usual single-particle shifts for each particle species. That is, if states $i_1,f_1$ belong to particles of type 1, and states $i_2,f_2$ belong to those of type 2, the total coordinate shift becomes
\begin{align}
      \d \rr^{f_1f_2}_{i_1i_2}=\d\rr^{(1)}_{f_1,i_1}+   \d\rr^{(2)}_{f_2,i_2},
\end{align}
where $\d\rr^{(1,2)}_{f,i}$ are the usual single-particle shifts for particles of types 1 and 2.

\textit{Coordinate shifts for Weyl fermions --} To illustrate the obtained result, we choose a model of fermions often considered in literature: that of a single species of Weyl fermions in free space, with local interaction. The free-fermion Hamiltonian for this model is
\begin{align}\label{eq:Weylhamiltonian}
  \hat H_0=\sum_{\pp}a^\dagger_{\pp,s}(v\pp\bm \sigma)_{ss'}a_{\pp,s'},
\end{align}
where $v$ is the speed of fermions, $\bm \sigma$ is a vector of Pauli matrices, and $s,s'$ are the spin indices.

To apply Eq.~\eqref{eq:shiftfinal} to collisions of fermions described by Hamiltonian~\eqref{eq:Weylhamiltonian}, we specialize to the ``particle'' band, $\varepsilon_\pp=vp$, in which the spinor describing a particle with momentum $\pp$ is given by
\begin{align}\label{eq:Wyelspinor}
  |u_\pp\rangle=\left(
  \begin{array}{c}\cos\frac{\theta_\pp}2\\ \sin \frac{\theta_\pp}2 e^{i\phi_\pp} \end{array}
  \right),
\end{align}
where $\theta_\pp$ and $\phi_\pp$ are the standard spherical angles of vector $\pp$. Written in the basis of states~\eqref{eq:Wyelspinor}, the interaction amplitude, Eq.~\eqref{eq:Hamiltonian}, for the particle band becomes
\begin{align}
  v_{\kk_1\kk_2;\pp_1\pp_2}=\frac{\lambda}{{\cal V}} \langle u_{\kk_1}|u_{\pp_2} \rangle \langle u_{\kk_2}|u_{\pp_1} \rangle \d_{\kk_1+\kk_2,\pp_1+\pp_2},
\end{align}
where $\lambda$ is the strength of the local interaction, and $\cal V$ is the volume of the crystal.

\begin{widetext}
The interaction matrix element that defines the two particle coordinate shift, Eq.~\eqref{eq:shiftfinal}, is given by
\begin{align}
  V_{\kk_1\kk_2;\pp_1\pp_2}=\frac{\lambda}{V} (\langle u_{\kk_1}|u_{\pp_1} \rangle \langle u_{\kk_2}|u_{\pp_2} \rangle-
  \langle u_{\kk_1}|u_{\pp_2} \rangle \langle u_{\kk_2}|u_{\pp_1} \rangle) \d_{\kk_1+\kk_2,\pp_1+\pp_2},
\end{align}
and has proper antisymmetry with respect to the  interchange of initial or final particles.

Defining a unit vector $\ee_\pp=\pp/p$, we obtain the coordinate shift for the $(\pp_1\pp_2)\to(\kk_1\kk_2)$ collision:

\begin{align}\label{eq:Weylshiftfinal}
   \d \rr^{\kk_1\kk_2}_{\pp_1\pp_2}=-\frac{v}2 \left(\frac{1}{\ve_{k_1}}-\frac{1}{\ve_{k_2}}\right)\frac{\ee_{\kk_1}\times\ee_{\kk_2}}{1-\ee_{\kk_1}\cdot\ee_{\kk_2}}
   +\frac{v}2 \left(\frac{1}{\ve_{p_1}}-\frac{1}{\ve_{p_2}}\right)\frac{\ee_{\pp_1}\times\ee_{\pp_2}}{1-\ee_{\pp_1}\cdot\ee_{\pp_2}}.
 \end{align}
\end{widetext}
We would like to emphasize that Eq.~\eqref{eq:Weylshiftfinal} gives the total coordinate shift for the system of the two scattering Weyl fermions. One must note that in Ref.~\onlinecite{Chen2014}, \emph{individual} coordinate shifts were defined for Weyl fermions interacting via a local interaction in free space in a reference frame where the collision is head-on. Using appropriate  Lorentz transformations, one can then consider the collision in any reference frame.  To see the relation between the results of this paper and those of Ref.~\onlinecite{Chen2014}, we note that the sum of the individual shifts for scattered particles defined in Ref.~\onlinecite{Chen2014} corresponds to the first term in Eq.~\eqref{eq:Weylshiftfinal} above, while the second term is trivially zero due to the collision being head-on ($\bm e_\pp\times\bm e_{\pp'}=0$). Further, for a head-on collision of particles with zero-range interaction and zero impact parameter, as in Ref.~\onlinecite{Chen2014}, the individual shifts can be counted from the collision point, and thus are physically well defined.

All these considerations break down for a collision in a crystal, and for a finite-range interparticle interaction potential. Indeed, the crystal represents a preferred reference frame, in which the Hamiltonian~\eqref{eq:Hamiltonian} of the colliding particles is given, eliminating the Lorentz invariance; any finite-range interaction potential allows non-zero impact parameters even for a head-on arrangement of the colliding particles momenta, making individual shifts undefined due to indistinguishability of particles. Therefore, in the realm of solid state physics, Eq.~\eqref{eq:shiftfinal} appears to be the only statement one can make regarding the net shift of the system of two colliding particles. The presence of a crystal also makes the collision effectively a three-body one, one assumed infinitely massive; this makes sure that the net collisional shift of the two-electron system's center of mass does not violate basic physical principles.

\textit{Application to the anomalous Hall effect --} Two-particle coordinate shifts have two-fold effect on electron kinetics, much in the same way it happens with single-particle shifts upon impurity scattering: they modify the expression for the electric current, and lead to the appearance of an additional generation term in the Boltzmann equation, stemming from the collision integral not being nullified by the equilibrium distribution function in the presence of an external electric field~\cite{SinitsynReview}.

The contribution to the electric current associated with the accumulation of two-particle shift events can be obtained from the Fermi golden rule considerations. Indeed, consider $\ell_1\ell_2\to\ell_3\ell_4$ scattering events, which happen at the rate of $W^{\ell_3\ell_4}_{\ell_1\ell_2}$. Since the pair of electrons gets shifted by $\d \rr^{\ell_3\ell_4}_{\ell_1\ell_2}$ in such events, they contribute $e W^{\ell_3\ell_4}_{\ell_1\ell_2}\d \rr^{\ell_3\ell_4}_{\ell_1\ell_2}$ to the electric current. Summing over all initial and final states, and including standard factors associated with the Fermi statistics, we obtain the following generalization of the ``shift accumulation'' current, $ \jj^{\textrm{shift}}$, to the present situation:
\begin{align}\label{eq:sjaccumulation}
  \jj^{\textrm{shift}}=\frac14e\sum_{\ell_1\ell_2\ell_3\ell_4}W^{\ell_3\ell_4}_{\ell_1\ell_2}\d \rr^{\ell_3\ell_4}_{\ell_1\ell_2}(1-f_{\ell_3})(1-f_{\ell_4})f_{\ell_1}f_{\ell_2}.
\end{align}
The factor of $1/4$ removes double-counting of initial and final state pairs, which occurs due to indistinguishability of particles. Note also that the expression for the shift accumulation current does not include the reverse processes, $\ell_3\ell_4\to\ell_1\ell_2$: due to the summation over all four of initial and final momenta in Eq.~\eqref{eq:sjaccumulation}, inclusion of such processes would constitute double-counting. The antisymmetry of $\d \rr^{\ell_3\ell_4}_{\ell_1\ell_2}$ with respect to the interchange of initial and final states, the scattering rate $W^{\ell_3\ell_4}_{\ell_1\ell_2}$ being symmetric, guarantees that current Eq.~\eqref{eq:sjaccumulation} vanishes in equilibrium.

For linear transport, one can directly relate the shift accumulation current~\eqref{eq:sjaccumulation} to the deviation of the distribution function from the equilibrium one, $f_0(\ve_\ell)$, defined via
\begin{align}\label{eq:linF}
  f_{\ell}=f_0(\ve_{\ell})-\phi_\ell\frac{\partial f_0(\ve_\ell)}{\partial\ve_\ell}.
  \end{align}
Nothing that
\begin{align}
  \frac{\partial f_0(\ve_\ell)}{\partial \ve_\ell}=-\frac{1}{T}f_0(\ve_\ell)(1-f_0(\ve_\ell)),\nonumber
\end{align}
we obtain the final expression for the shift accumulation current:
\begin{align}\label{eq:sjaccumulationlinearized}
  \jj^{\textrm{shift}}=\frac {e}{2T}\sum_{\ell_1\ell_2\ell_3\ell_4}W^{\ell_3\ell_4}_{\ell_1\ell_2}
  \d \rr^{\ell_3\ell_4}_{\ell_1\ell_2}(1-f_{\ell_3})(1-f_{\ell_4})f_{\ell_1}f_{\ell_2}\phi_{\ell_1}.
\end{align}

We now turn to the additional generation term in the Boltzmann equation, which stems from the fact that the work done by the electric field due to the shift during the collision must be taken into account in the energy conservation. This fact is taken into account by modifying the energy-conserving $\d$-function in the collision integral:
\begin{align}
 \d(\e_{\ell_3}+\e_{\ell_4}-\e_{\ell_1}-\e_{\ell_2})\to \d(\e_{\ell_3}+\e_{\ell_4}-\e_{\ell_1}-\e_{\ell_2}-e\EE\d \rr^{\ell_3\ell_4}_{\ell_1\ell_2})\nonumber.
\end{align}
As in the single-particle case~\cite{Sinitsyn2006}, this means that in the presence of the coordinate shifts the collision integral is not nullified by the Fermi-Dirac distribution function. Expanded to the linear order in the external electric field $\EE$, the electron-electron collision integral then contains an effective generation term, $-e\EE \bm g_\ell$ with $\bm g_\ell$, given by
\begin{align}\label{eq:generationG}
  \bm g_{\ell_1}&=\frac{1}{2T}\sum_{\ell_2\ell_3\ell_4}W^{\ell_3\ell_4}_{\ell_1\ell_2}
  \d \rr^{\ell_3\ell_4}_{\ell_1\ell_2}(1-f_{\ell_3})(1-f_{\ell_4})f_{\ell_1}f_{\ell_2},
\end{align}
which can be combined with the usual electric drive term in the left hand side of the kinetic equation for the stationary non-equilibrium state:
\begin{align}\label{eq:kineq}
  e\EE\left(\bm v_\ell\frac{\partial f_0(\ve_\ell)}{\partial\ve_\ell}+\bm g_\ell\right)=I_{st}(\phi_\ell).
\end{align}
In this equation, $I_{st}(\phi_\ell)$ is the linearized collision integral, which in general must contain electron-impurity and electron-phonon contributions, necessary to reach a steady state in the presence of an external electric field.

The kinetic equation~\eqref{eq:kineq} makes it clear that the deviation from the equilibrium is created by both acceleration by $\EE$, as well as two-particle coordinate shifts. The corresponding changes in the distribution function, Eq.~\eqref{eq:linF}, are given by
\begin{align}
  \phi^{\bm v}&=I_{st}^{-1}\left(e\EE\bm v_\ell\frac{\partial f_0(\ve_\ell)}{\partial\ve_\ell}\right),\nonumber\\
   \phi^{\bm g}&=I_{st}^{-1}\left(e\EE\bm g_\ell\right).
\end{align}
The assumed presence of the electron-impurity and electron-phonon contributions to the collision integral ensures that the inverse operator, $I_{st}^{-1}$, is defined for any generation term in the left hand side of the kinetic equation.

Finally, the expression for the current contribution due to the two-particle coordinate shifts is given by
\begin{align}
 \jj^{\textrm{AHE}}=\jj^{\textrm{shift}}+\jj^{\textrm{ballistic}},
\end{align}
where
\begin{align}\label{eq:ballisticj}
\jj^{\textrm{ballistic}}=-\frac{e}{\cal V}\sum_{\ell}\bm v_\ell\frac{\partial f_0(\ve_\ell)}{\partial\ve_\ell}\phi^{\bm g}_\ell,
\end{align}
and
\begin{align}\label{eq:shiftj}
\jj^{\textrm{shift}}=\frac{e}{\cal V}\sum_{\ell}\bm g_\ell \phi^{\bm v}_\ell.
\end{align}
We used Eqs.~\eqref{eq:sjaccumulationlinearized} and \eqref{eq:generationG} to write down the above expression for $\jj^{\textrm{shift}}$. One can show that the conductivity tensor that defines the linear relationship between $\jj^{\textrm{AHE}}$ and the electric field $\EE$ is indeed antisymmetric, as appropriate for the anomalous Hall effect.
Expressions~\eqref{eq:generationG}, \eqref{eq:ballisticj}, and \eqref{eq:shiftj} are one of the central results of this work.

An illustration of a solution of kinetic equation~\eqref{eq:kineq} for the two-dimensional massive Dirac model will be presented in a forthcoming publication~\cite{footnoteAHE}. Here we only mention that the generation term due to two-particle coordinate shifts, Eq.~\eqref{eq:generationG}, conserves the particle number, the total momentum, and the total energy of the electronic liquid. This can be verified in the usual way~\cite{LL10} by using Eq.~\eqref{eq:generationG}, antisymmetry of $\d \rr^{\ell_3\ell_4}_{\ell_1\ell_2}$, and symmetry of the scattering rate $W^{\ell_3\ell_4}_{\ell_1\ell_2}$with respect to the interchange of initial and final pairs states. This fact can be shown to ensure that the present mechanism does not contribute to the transport coefficients for purely parabolic dispersion, as expected.

\textit{Conclusions -- } In this paper, we have introduced the notion of the two-particle collisional coordinate shift, occurring in electron-electron collisions in crystals. The expression for the shift, Eq.~\eqref{eq:shiftfinal}, is valid for arbitrary band structure, and arbitrary interparticle interaction potential, which can be treated in Born approximation -- a typical assumption in treatment of carrier collisions in semiconductors. We further showed that the two-particle shifts make a contribution to the anomalous Hall effect in the hydrodynamic regime.

While Eq.~\eqref{eq:shiftfinal} was derived for particle collisions in vacuum, its validity is more general. In particular, the expression is valid also in many-particle systems, in which all the other particles is treated as a `refractive medium' for the two colliding particles, which, in RPA-class of approximations, leads to the dependence of the effective interaction matrix elements on the transferred energy. The effective interaction, however, has to be treated in the Born approximation. The RPA-type renormalizations of the effective interaction can bring in a true many-body physics into the problem of anomalous transport in hydrodynamic electronic systems. Further investigation along this direction, as well as inclusion of the Fermi liquid effects, represents an interesting and open theoretical problem.

We acknowledge very useful discussions with Anton Andreev, Eugene Mishchenko, Dam Thanh Son, Oleg Starykh, and Mikhail Stephanov.

\bibliography{hAHE_references}

\begin{thebibliography}{43}
\bibitem[{foo()}]{footnoteAHE}
\bibinfo{note}{The details of the calculation will be reported in a forthcoming
  publication by the author.}
\end{thebibliography}
\bibliographystyle{apsrev}

\end{document}